\documentclass[12pt,preprint]{emulateapj}

\newcommand{\eps}{\varepsilon}

\begin{document}

\title{Radiative MHD simulation of sunspot structure}
\shorttitle{Radiative MHD simulation of sunspot structure}
\shortauthors{Rempel, Sch\"ussler \& Kn\"olker}

\author{M. Rempel\altaffilmark{1}, M. Sch\"ussler\altaffilmark{2}  
  and M. Kn\"olker\altaffilmark{1}}

\email{rempel@hao.ucar.edu}

\altaffiltext{1}{High Altitude Observatory,
    NCAR, P.O. Box 3000, Boulder, Colorado 80307, USA}
\altaffiltext{2}{Max-Planck-Institut f\"ur Sonnensystemforschung,
    Max-Planck-Str. 2, 37191 Katlenburg-Lindau, Germany}

\begin{abstract}
  Results of a 3D MHD simulation of a sunspot with a photospheric size
  of about 20~Mm are presented. The simulation has been carried out with
  the MURaM code, which includes a realistic equation of state with
  partial ionization and radiative transfer along many ray
  directions. The largely relaxed state of the sunspot shows a division 
  in a central dark umbral region with bright dots and a penumbra showing
  bright filaments of about $2$ to $3$ Mm length with central dark lanes. 
  By a process similar to the formation of umbral dots, the penumbral
  filaments result from magneto-convection in the form of upflow plumes,
  which become elongated by the presence of an inclined magnetic field:
  the upflow is deflected in the outward direction while the magnetic
  field is weakened and becomes almost horizontal in the upper part of the 
  plume near the level of optical depth unity. A dark lane forms owing to 
  the piling up of matter near the cusp-shaped top of the rising plume
  that leads to an upward bulging of the surfaces
  of constant optical depth. The simulated penumbral structure
  corresponds well to the observationally inferred interlocking-comb
  structure of the magnetic field with Evershed outflows along
  dark-laned filaments with nearly horizontal magnetic field and
  overturning perpendicular (`twisting') motion, which are embedded in a
  background of stronger and less inclined field. Photospheric spectral
  lines are formed at the very top and somewhat above the upflow plumes,
  so that they do not fully sense the strong flow as well as the large
  field inclination and significant field strength reduction in the
  upper part of the plume structures.
\end{abstract}




\section{Introduction}

The physical understanding of sunspot structure has been hampered for
decades by 1) insufficient resolution of the fine structure by
observations, 2) lack of information about the layers below the visible
surface, and 3) insufficient computational power to perform ab-initio 3D
MHD simulation of a full sunspot including the surrounding
granulation. Recently, we have seen considerable progress on all three
of these fronts: 1) adaptive optics, image selection and reconstruction
at ground-based telescopes and the advent of spectro-polarimetry in the
visible from space with the {\em Hinode} satellite have led to a wealth
of new information about the fine structure of sunspot umbrae and
penumbrae \citep[e.g.][]{Bharti:etal:2007, Langhans:etal:2007,
Ichimoto:etal:2007, Riethmueller:etal:2008, Rimmele:Marino:2006} 2)
local helioseismology has started to probe the sub-surface structure of
sunspots \citep[e.g.,][]{Cameron:etal:2008}, and 3) the ever-increasing
computational power of parallel computers have made ab-initio
simulations of full sunspots come into reach.

While \citet{Cameron:etal:2007b} simulated solar pores of up to about
3~Mm diameter and did not find indications for the development of a
penumbral structure, the first attempt to simulate a sunspot together
with the surrounding granulation is due to
\citet{Heinemann:etal:2007}. They considered a rectangular section of a
(slab-like) small sunspot of about 4~Mm diameter. The main result of
this simulation is the formation of filamentary structures in the outer
part of the spot, various properties of which (such as dark cores,
outflows, and strongly inclined magnetic field) are consistent with
observational results.  However, the filaments found by
\citet{Heinemann:etal:2007} are much shorter than the typical lengths of
real penumbral filaments and the overall extension of the simulated
penumbra is very small.

Here we report about results of a sunspot simulation with the {\em
MURaM} code \citep{Voegler:etal:2005}. The simulated sunspot has a
total diameter of about 20~Mm, shared about equally by umbra and
penumbra. 

\section{Simulation setup}

\begin{figure*}
  \centering 
  \resizebox{\hsize}{!}{\includegraphics{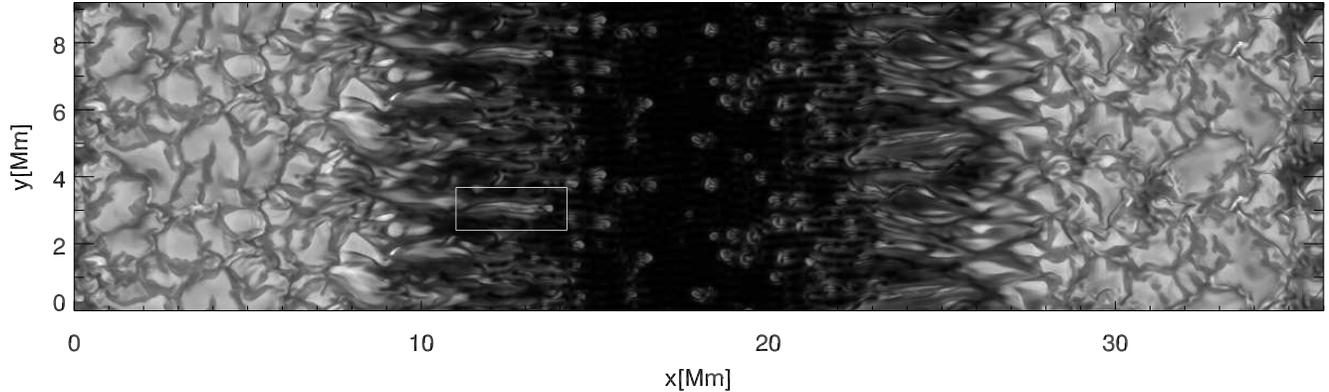}}
  \caption{Continuum intensity image at 630~nm of the simulated sunspot
    and its environment (doubled in the $y$-direction). The bright
    umbral dots and penumbral filaments have peak intensities between
    40\% and 90\% of the average value outside the spot. The penumbral
    filaments reach lengths of 2--3 Mm. The white frame indicates the
    filament studied in detail in Sec.~3.2.}
  \label{fig:intensity_global}
\end{figure*}

\begin{figure*}
  \centering 
  \resizebox{\hsize}{!}{\includegraphics{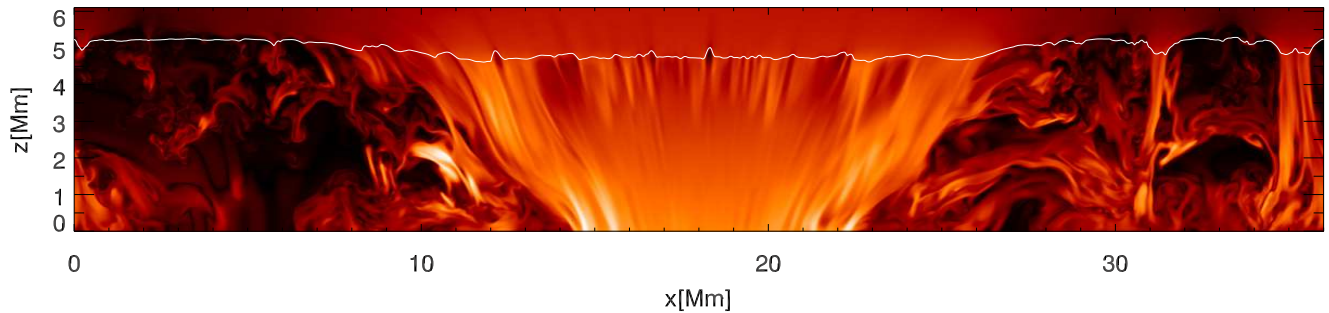}}
  \caption{Vertical structure of the simulated sunspot. Shown is a color
  coding of the square root of the field strength on a cut through the
  simulation in $x$-$z$ plane at $y=3.84\,$Mm. White color corresponds
  to the maximum field strength of 9$\,$kG, black to zero field. The
  whitish line indicates the height of the level $\tau_{630}=1.$ The
  average Wilson depression of the spot umbra is about 450~km. Separated
  from the main sunspot, two pore-like structures have formed at
  $x\simeq31\,$Mm and $x\simeq35\,$Mm, respectively.}
  \label{fig:bb_vert_global}
\end{figure*}

\begin{figure}
  \centering 
  \resizebox{\hsize}{!}{\includegraphics{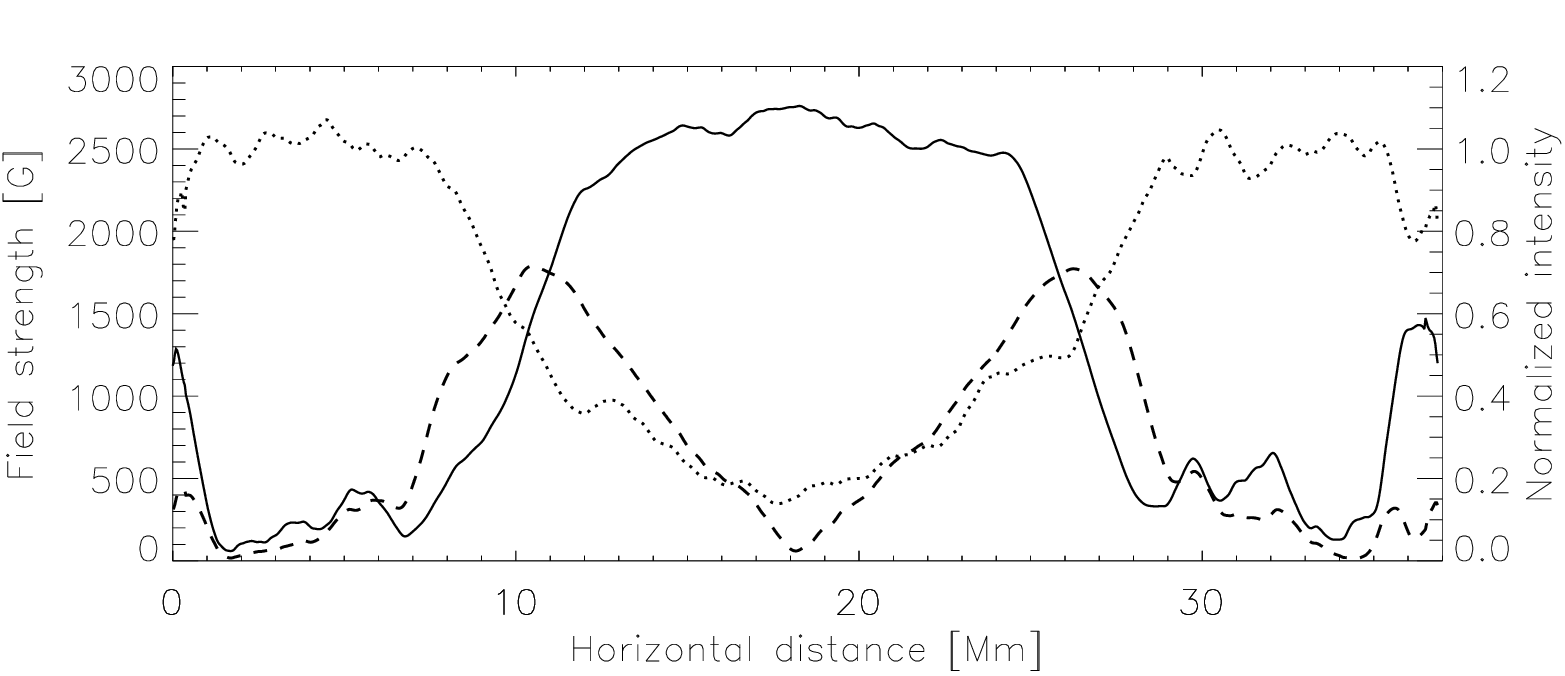}}
  \resizebox{\hsize}{!}{\includegraphics{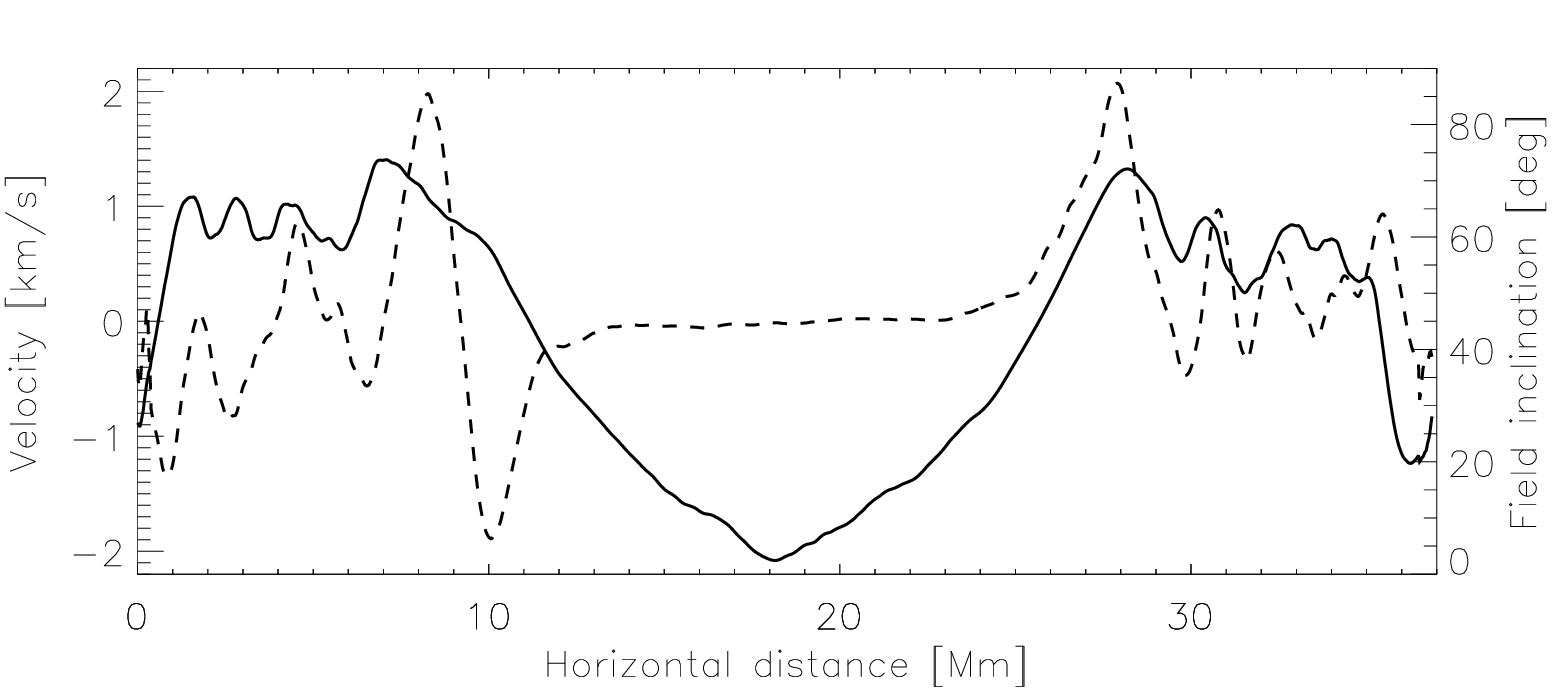}}
  \caption{Large-scale structure of the simulated sunspot: quantities
  averaged over the horizontal $y$-direction as functions of $x$.  {\em
  Upper panel:} vertical magnetic field, $B_z$, at $\tau_{630}=0.1$
  (solid), modulus of the horizontal field component, $|B_x|$, at
  $\tau_{630}=0.1$ (dashed), and normalized continuum intensity at 630~nm
  (dotted). {\em Lower panel:} magnetic field inclination with respect
  to the vertical at $\tau_{630}=0.1$ (solid) and horizontal velocity,
  $v_x$, at $\tau_{630}=1.$ (dashed).}
  \label{fig:means_global}
\end{figure}

The primary numerical challenges of performing a large-scale sunspot
simulation including umbra, penumbra and granulation are the significant
variations of $\beta=8\pi p/B^2$ and Alfv{\'e}n velocity encountered at a
given geometrical height.  While a very low value of $\beta$ primarily
imposes stability problems, very high Alfv{\'e}n velocities of more that
$1000$~km$\cdot$s$^{-1}$ above the umbra of a spot lead to unacceptably
small time steps for an explicit code. In order to cope with these
problems, we have modified the {\em MURaM} code as outlined in
Appendix~\ref{appendix}.

The basic concept of our simulation is similar to that of
\citet{Heinemann:etal:2007}: we consider a slender rectangular section
of a sunspot. However, our computational domain is considerably wider
and deeper than theirs, permitting us to simulate a much larger sunspot.
The rectangular box spans $36.864\,{\rm Mm}\times 4.608\,{\rm Mm}$ in
the horizontal $(x,y)$ directions and $6.144$ Mm in the vertical $(z)$
direction. The side boundaries are periodic. The top boundary is closed
for the flow and the magnetic field is matched to a potential field
above.  At the bottom, the boundary is open for the flow and the
magnetic field kept vertical \citep[for details,
see][]{Voegler:etal:2005}.  In regions of strong magnetic field
($B>1000$~G), the bottom boundary is closed to avoid outflow
instabilities in long runs. The radiative transfer is treated in the
grey approximation.

The simulation was started from a thermally relaxed non-magnetic run by
introducing a 2D (slab) field configuration in the $x$-$z$ plane with a
width of $5$ Mm and a strength of $10$ kG at the bottom of the box,
expanding to $15$ Mm at the top.  Using a moderate grid resolution of
$48\,{\rm km}\times48\,{\rm km}\times32\,{\rm km}$ we run the
simulation for about $12$ hours solar time. After a very dynamic
adjustment phase of about $2$ hours, elongated filaments with dark cores
of about $2$ to $3$ Mm length started forming in the periphery of the
umbra, their heads moving inward. After about $7.5$ hours of evolution we
restarted from a snapshot and increased the resolution to $32\,{\rm
km}\times32\,{\rm km}\times21.33\,{\rm km}$.

\section{Results}
\begin{figure}
  \centering 
  \resizebox{\hsize}{!}{\includegraphics{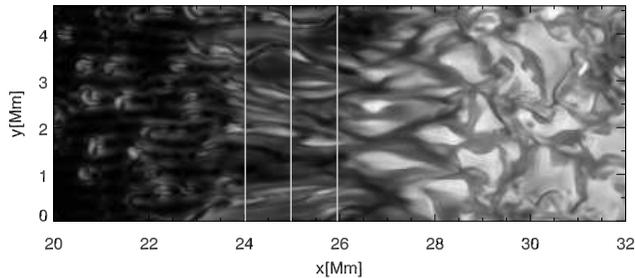}}
  \caption{Continuum intensity image showing details of penumbral
    filaments.  The vertical lines indicate the positions of the 
    vertical cuts presented in Fig.~\ref{fig:cuts_color}.}
  \label{fig:intensity_detail}
\end{figure}

\begin{figure}
  \centering 
  \resizebox{\hsize}{!}{\includegraphics{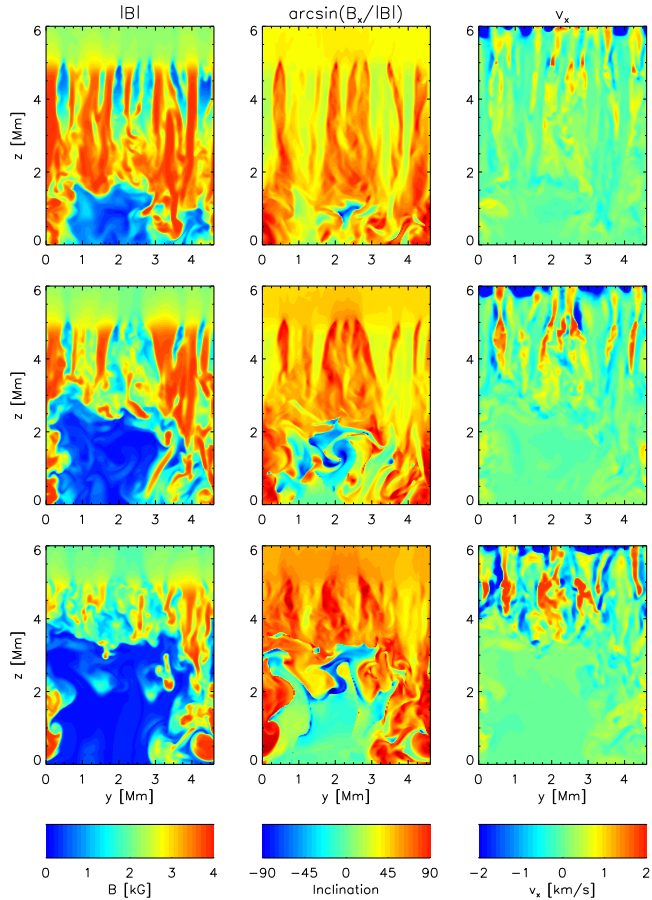}}
  \caption{Vertical cuts in the $y$-$z$ plane of magnetic field strength
    (left), inclination angle (middle) and horizontal velocity (right)
    for the positions indicated in Fig.~\ref{fig:intensity_detail} (the top
    row corresponding to the leftmost cut). The magnetic field strength is
    saturated at $4\,$kG, the velocity at $2\,$km$\cdot$s$^{-1}$. The
    filaments correspond to about $2\,$Mm deep channels of weaker and
    almost horizontal field and an outflow of a few km$\cdot$s$^{-1}$. 
  }
  \label{fig:cuts_color}
\end{figure}

\begin{figure}
  \centering 
  \resizebox{\hsize}{!}{\includegraphics{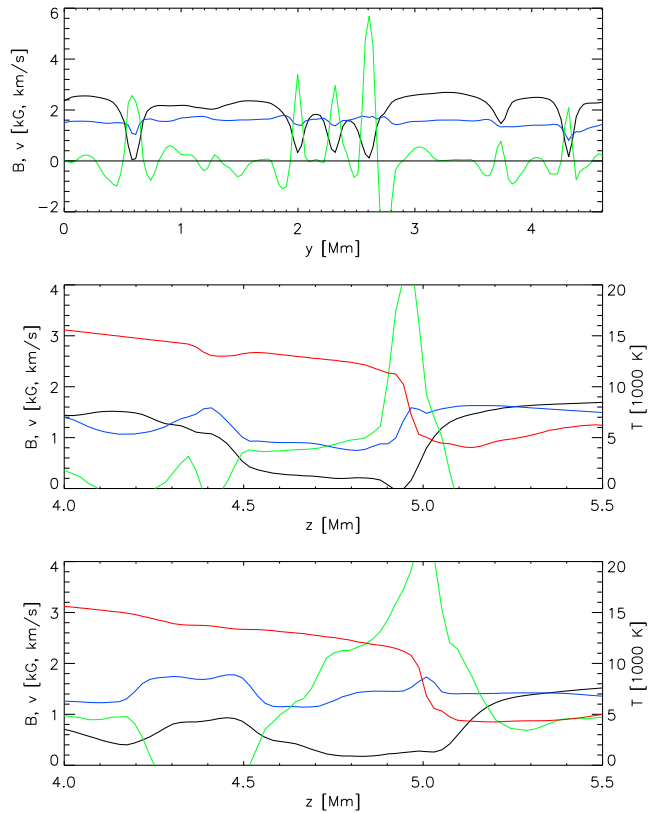}}
  \caption{Horizontal and vertical profiles of various physical
    quantities. {\em Top panel:} Horizontal cut along the middle line in
    Fig.~\ref{fig:intensity_detail} at height $z=5$ Mm, near the
    $\tau=1$ level in the filament. Shown are the vertical field
    component ($B_z$, black), the horizontal field component ($B_x$,
    blue), and the horizontal velocity component ($v_x$, green). The
    vertical field drops to very low values in the filaments, resulting
    in an almost horizontal field of $1-1.5\,$kG. Along the centers of
    the filaments, we have an outward directed horizontal flow. {\em Middle
    panel:} Vertical profiles of the same quantities through the center
    of the second filament from the left in the top panel. In addition,
    the red line shows the temperature profile. The horizontal outflow
    peaks close to the $\tau=1$ level, where the magnetic field has the
    largest inclination. {\em Bottom panel:} Vertical profiles at the
    center of the fourth filament from the left in the top panel.}
  \label{fig:profiles}
\end{figure}

\begin{figure*}
  \centering 
  \resizebox{\hsize}{!}{\includegraphics{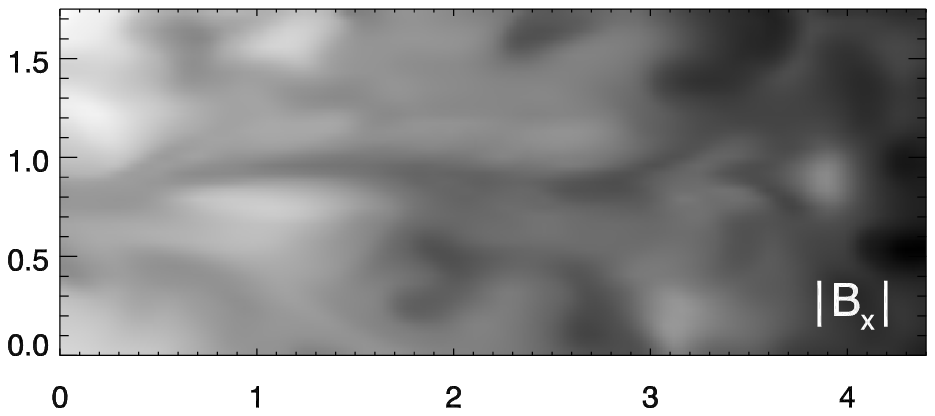}
	               {\includegraphics{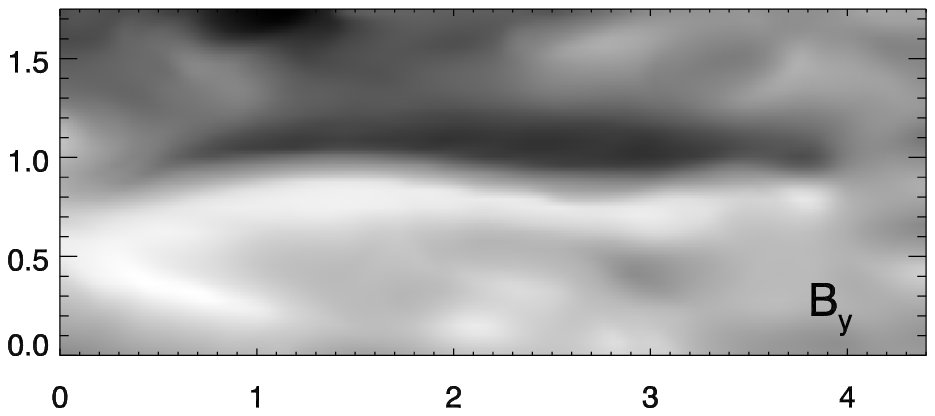}}}
  \resizebox{\hsize}{!}{\includegraphics{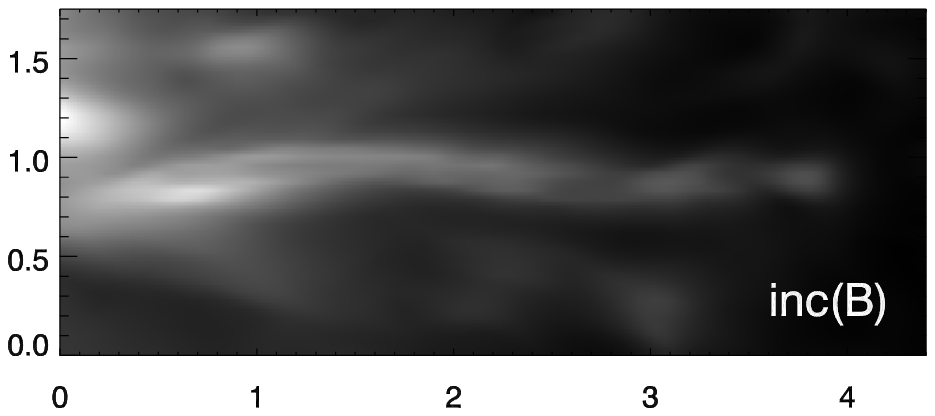}
	               {\includegraphics{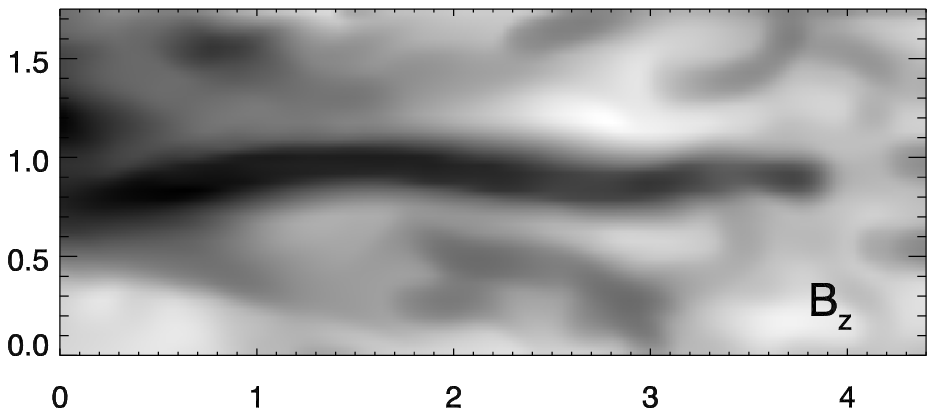}}}
  \resizebox{\hsize}{!}{\includegraphics{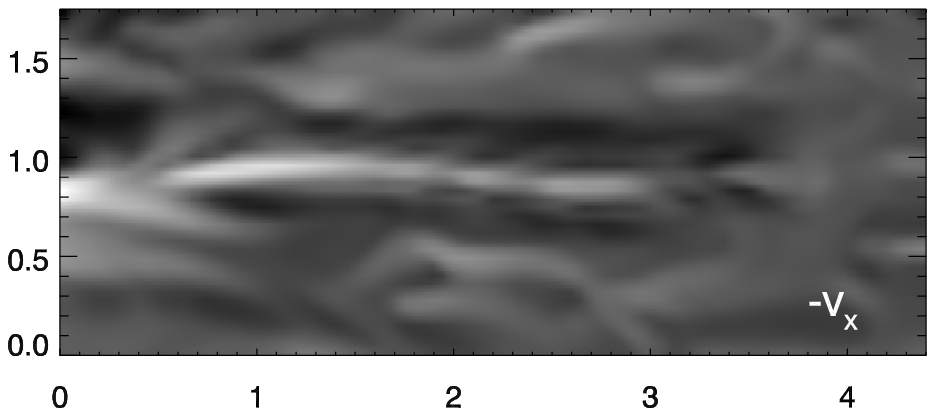}
	               {\includegraphics{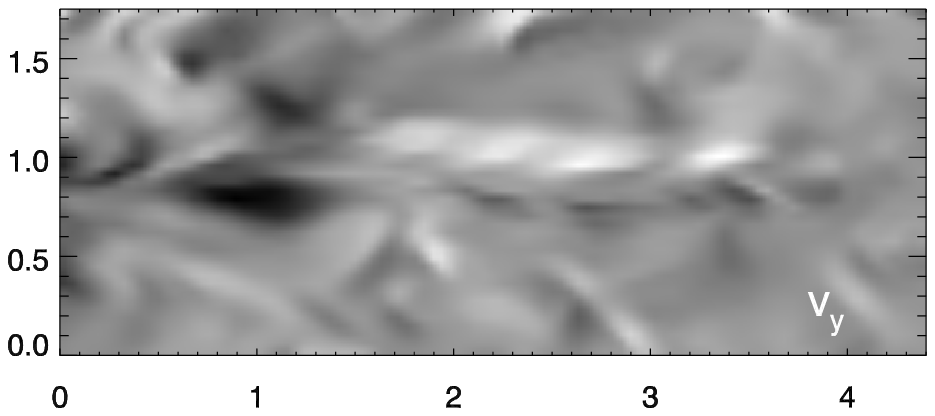}}}
  \resizebox{\hsize}{!}{\includegraphics{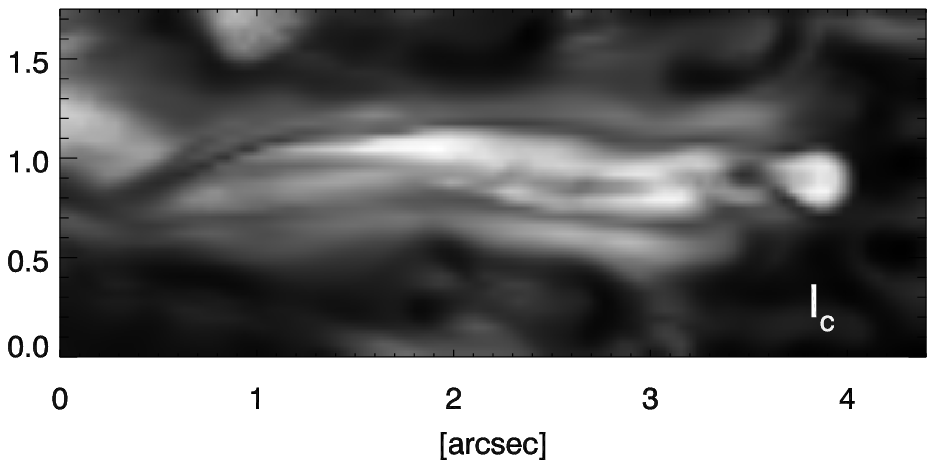}
	               {\includegraphics{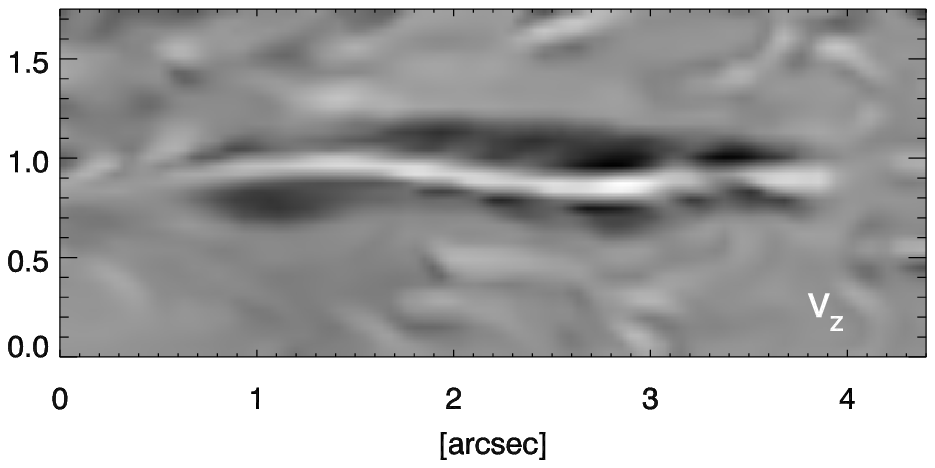}}}
  \caption{Grey-scale maps of physical quantities on the surface
    $\tau_{630}=0.1$ for the filament indicated by the rectangular box
    in Fig.~\ref{fig:intensity_global}. The ranges between minimum
    (black) and maximum (white) of the various quantities
    are: $|B_x|$: 670$\,$G ... 1970$\,$G; $B_y$: $-940\,$G ... 640$\,$G;
    $B_z$: 980$\,$G ... 3050$\,$G; $B$ inclination with respect to
    vertical: 17.5$\,$deg ... 61$\,$deg; $-v_x$:
    $-1.4\,$km$\cdot$s$^{-1}$ ...  3.3$\,$km$\cdot$s$^{-1}$; $v_y$:
    $-0.95\,$km$\cdot$s$^{-1}$ ...  0.66$\,$km$\cdot$s$^{-1}$; $v_z$:
    $-2.1\,$km$\cdot$s$^{-1}$ ...  1.5$\,$km$\cdot$s$^{-1}$; $I_{630}$:
    0.13 ... 1.02 of the average value outside the spot.}
  \label{fig:filament_hor}
\end{figure*}

\begin{figure}
  \centering 
  \resizebox{\hsize}{!}{\includegraphics{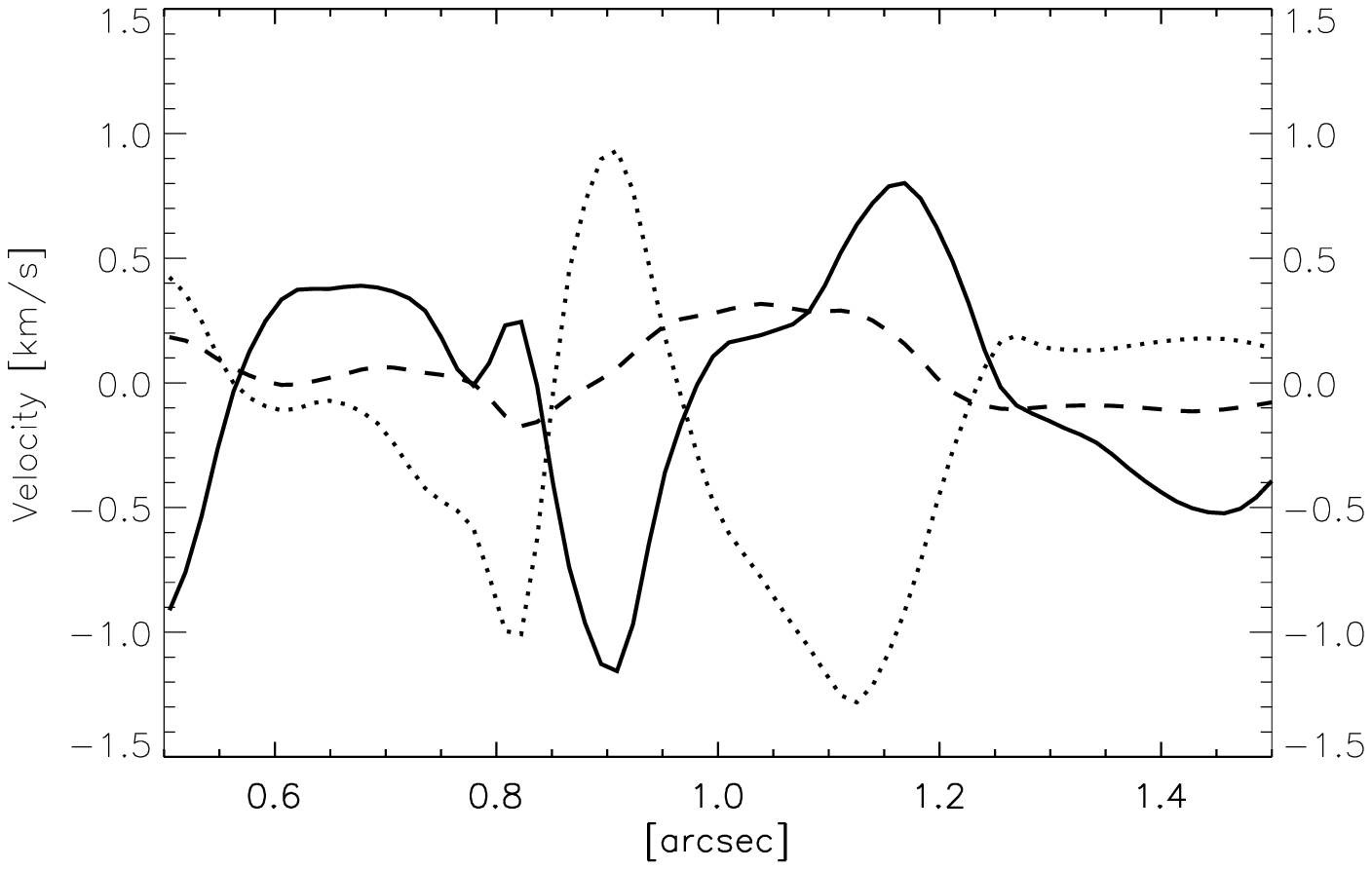}}
  \resizebox{\hsize}{!}{\includegraphics{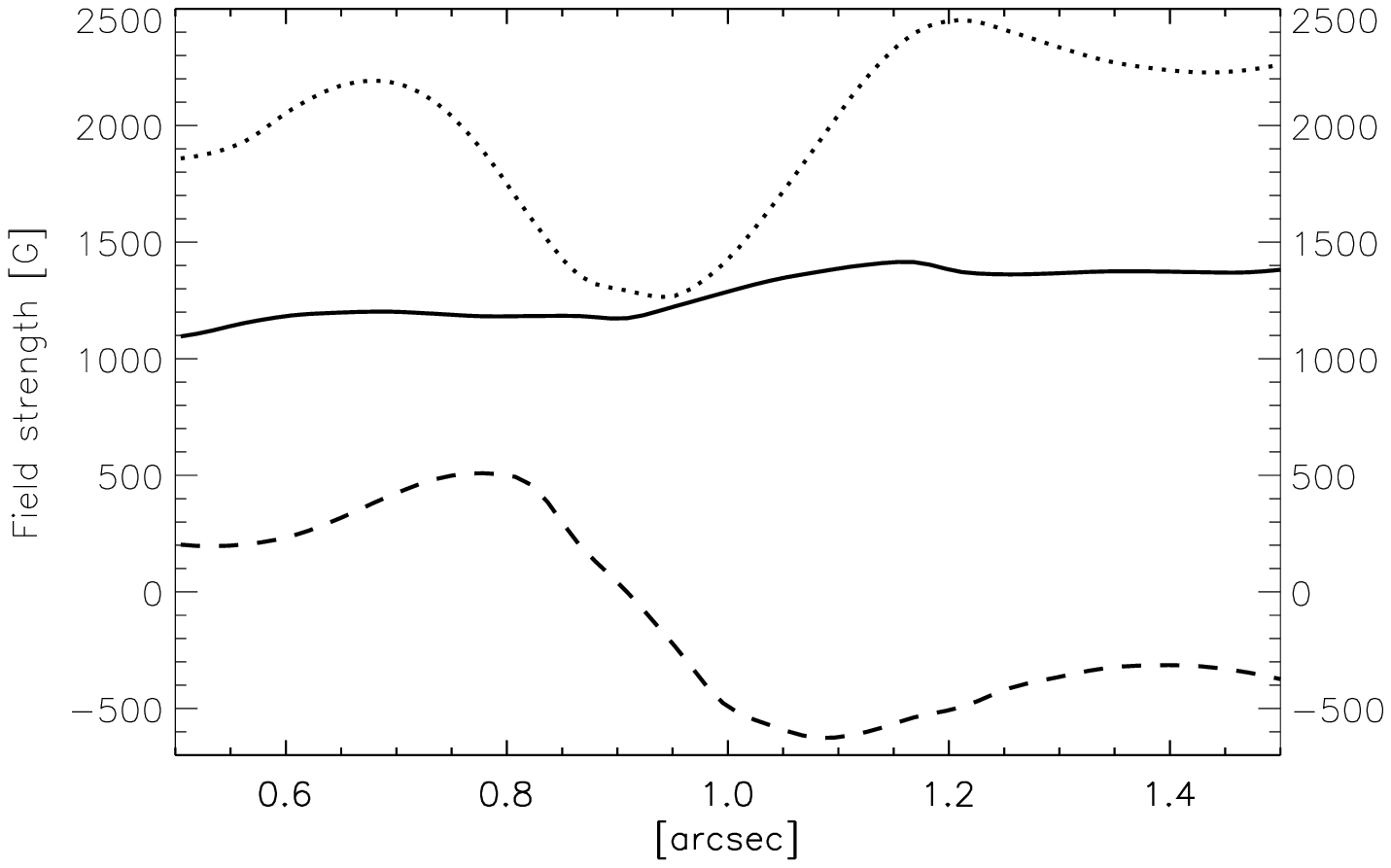}}
  \caption{Profiles of the magnetic field and velocity components at
           optical depth $\tau_{630}=0.1$ along a cut perpendicular to
           the filament shown in Fig.~\ref{fig:filament_hor} (cut in
           $y$-direction at $x=2\,$arcsec). {\em Upper panel:} $v_x$
           (solid), $v_y$ (dashed), $v_z$ (dotted). {\em Lower panel:}
           $B_x$ (solid), $B_y$ (dashed), $B_z$ (dotted).  }
  \label{fig:B_v_profiles}
\end{figure}

\begin{figure*}
  \centering 
  \resizebox{\hsize}{!}{
    \includegraphics{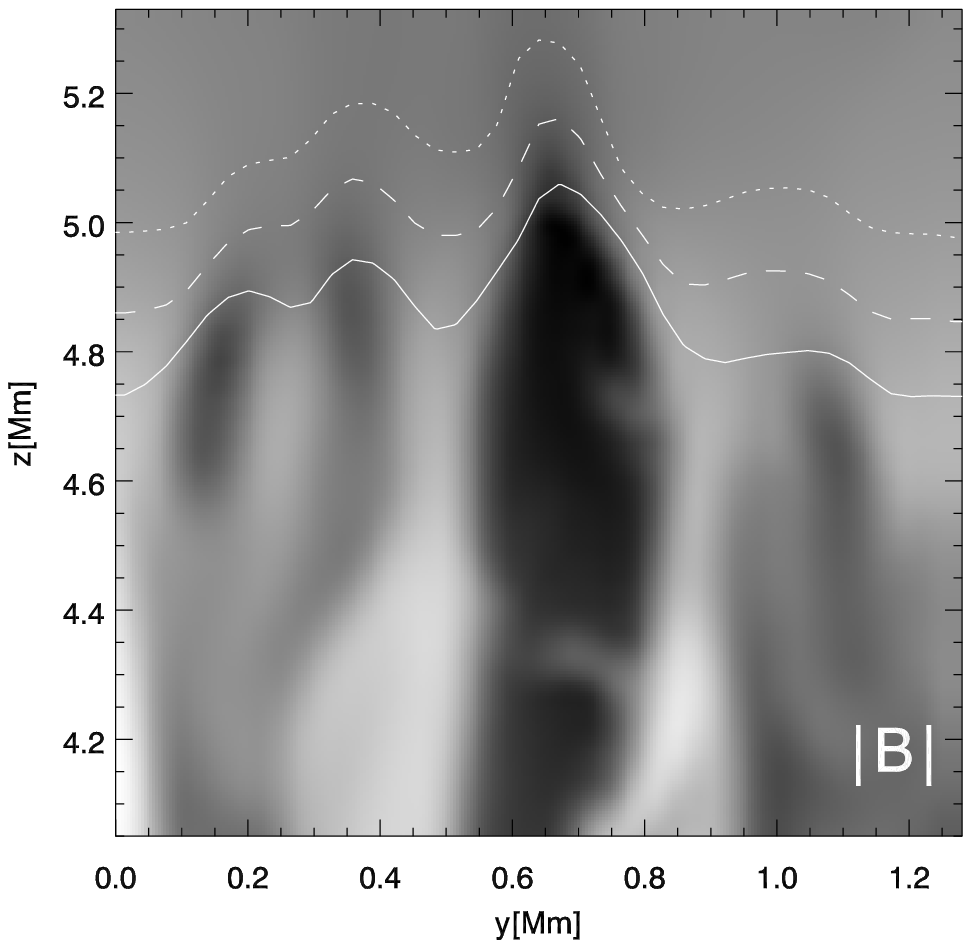}
    \includegraphics{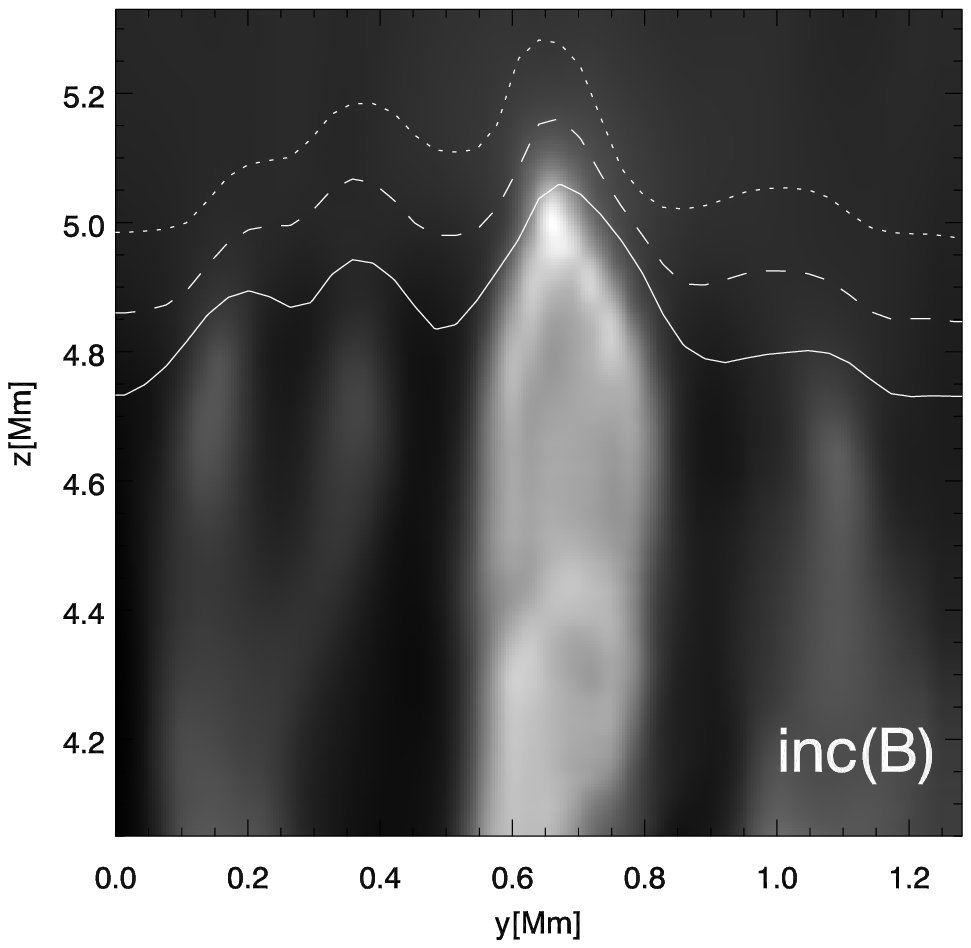}
    \includegraphics{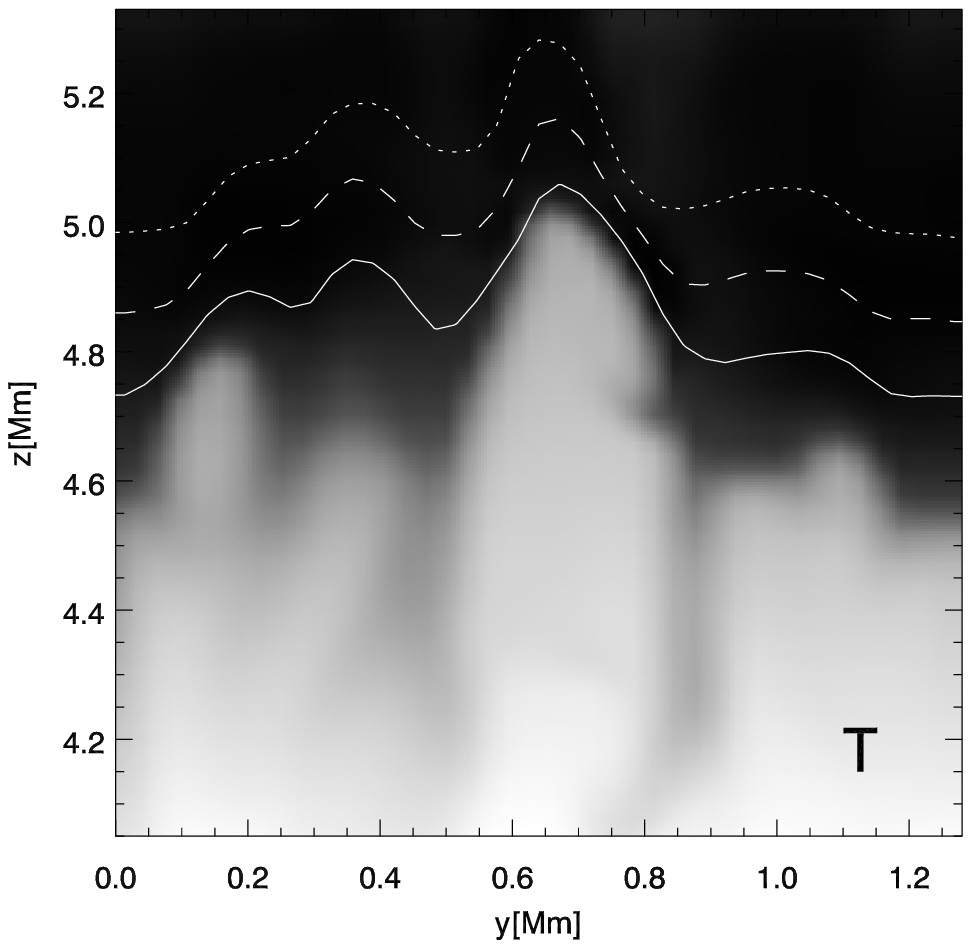}}
  \resizebox{\hsize}{!}{
    \includegraphics{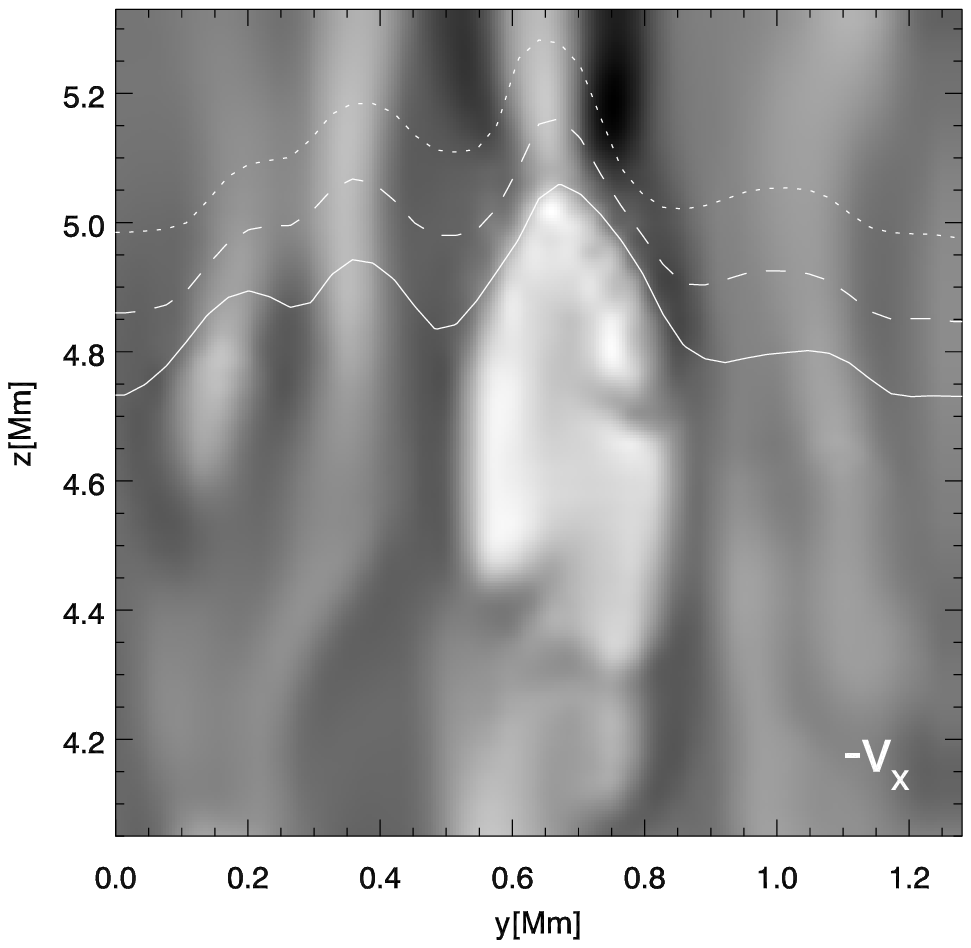} 
    \includegraphics{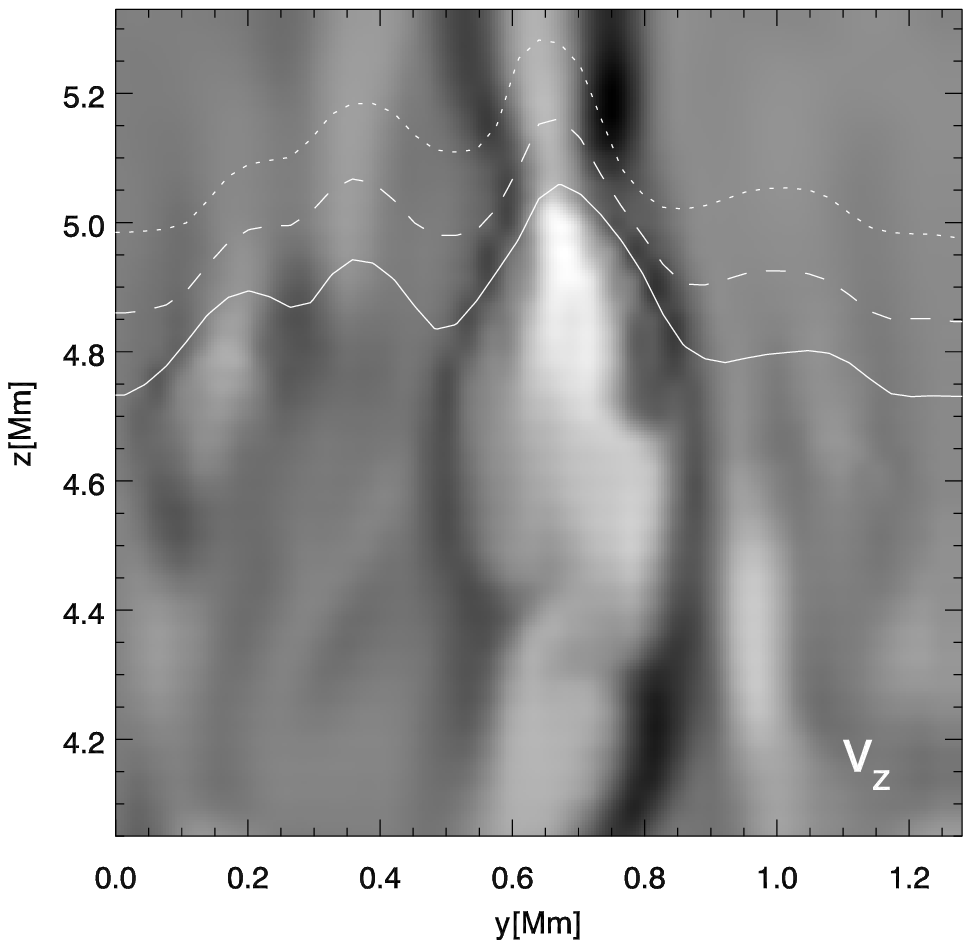}
    \includegraphics{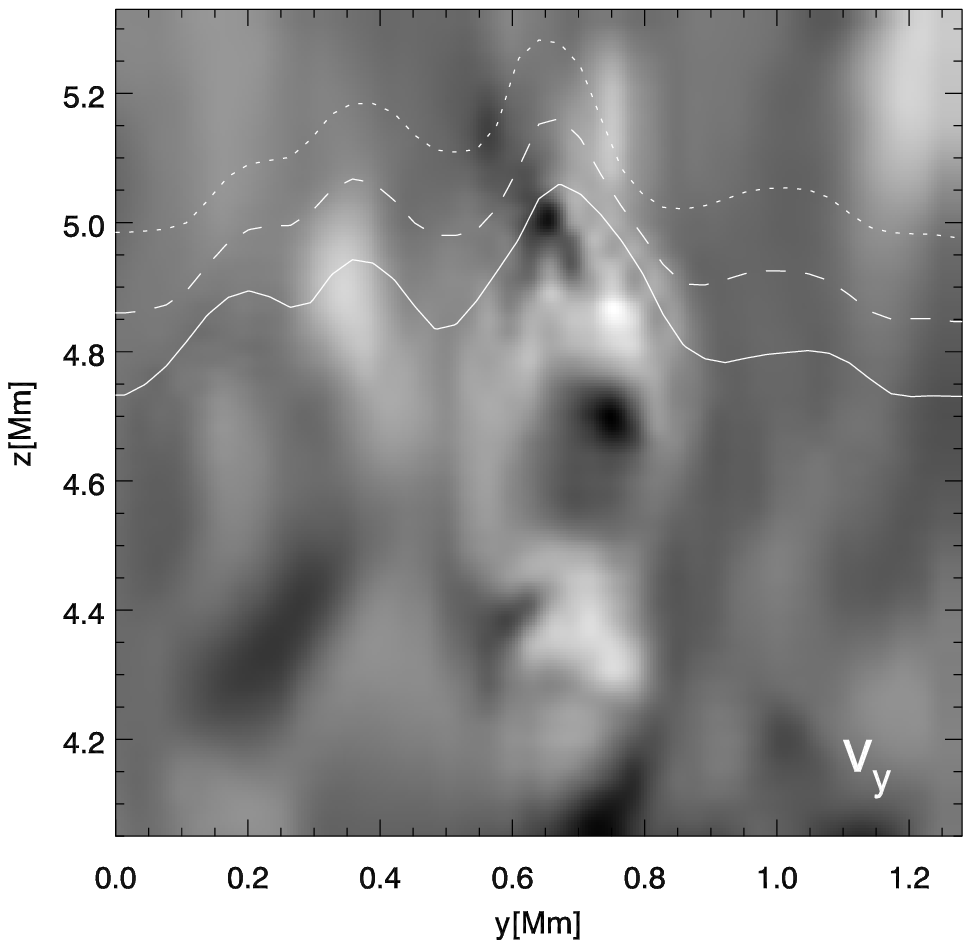}} 
  \caption{Vertical cuts at $x=2\,$arcsec perpendicular through the
   filament shown in Fig.~\ref{fig:filament_hor}. The vertical scale is
   the same as used in Figs.~\ref{fig:cuts_color} and 
   \ref{fig:profiles}, where zero height corresponds to the bottom of 
   the computational domain. The
   grey-scale ranges between minimum (black) and maximum (white) of the
   quantities shown are: $|B|$: 480$\,$G ... 4220$\,$G; $B$ inclination
   with respect to vertical: 20.1$\,$deg ... 90$\,$deg; $T$:
   $3890\,$K ... $15410\,$K; $-v_x$:
   $1.88\,$km$\cdot$s$^{-1}$ ...  2.09$\,$km$\cdot$s$^{-1}$; $v_y$:
   $-0.7\,$km$\cdot$s$^{-1}$ ...  0.8$\,$km$\cdot$s$^{-1}$; $v_z$:
   $-2.3\,$km$\cdot$s$^{-1}$ ...  2.2$\,$km$\cdot$s$^{-1}$ . 
   The white lines indicate the levels of 
   $\tau_{630}=1.$, 0.1, and 0.01, respectively.}
  \label{fig:filament_vertx}
\end{figure*}

\begin{figure*}
  \centering 
  \resizebox{\hsize}{!}{\includegraphics{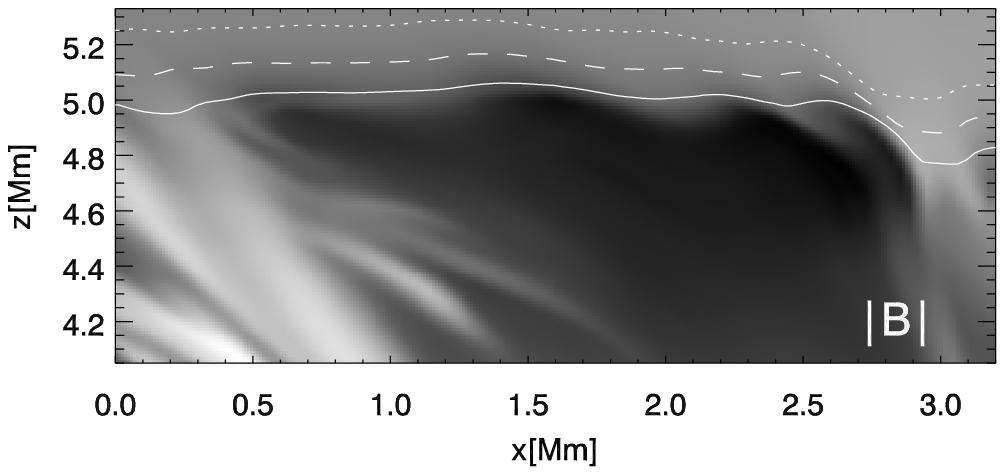}
	               {\includegraphics{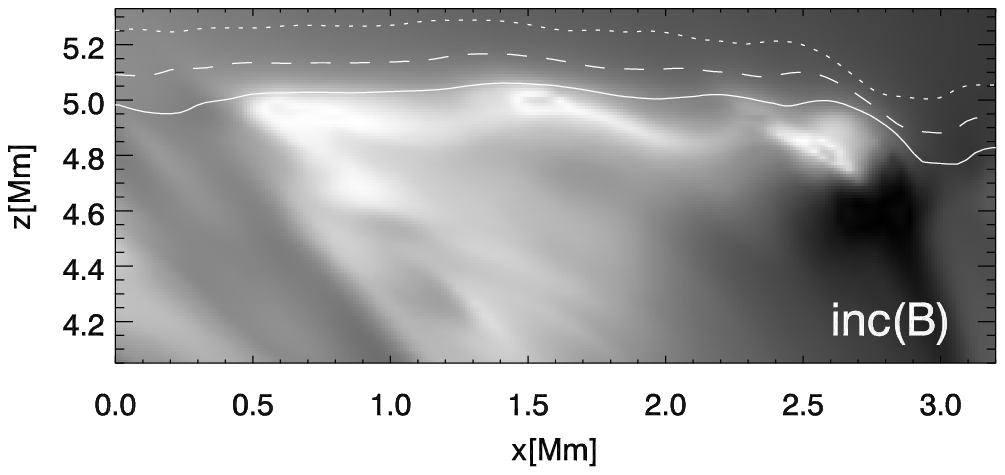}}}
  \resizebox{\hsize}{!}{\includegraphics{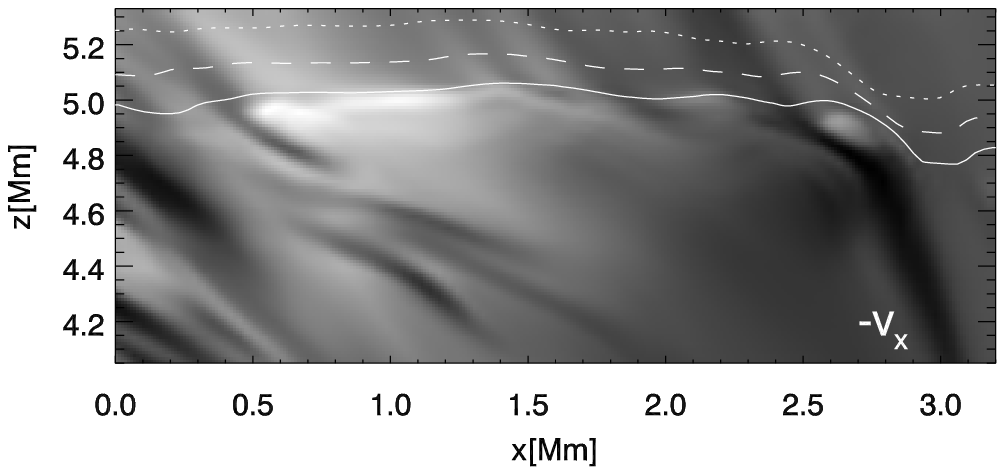}
	               {\includegraphics{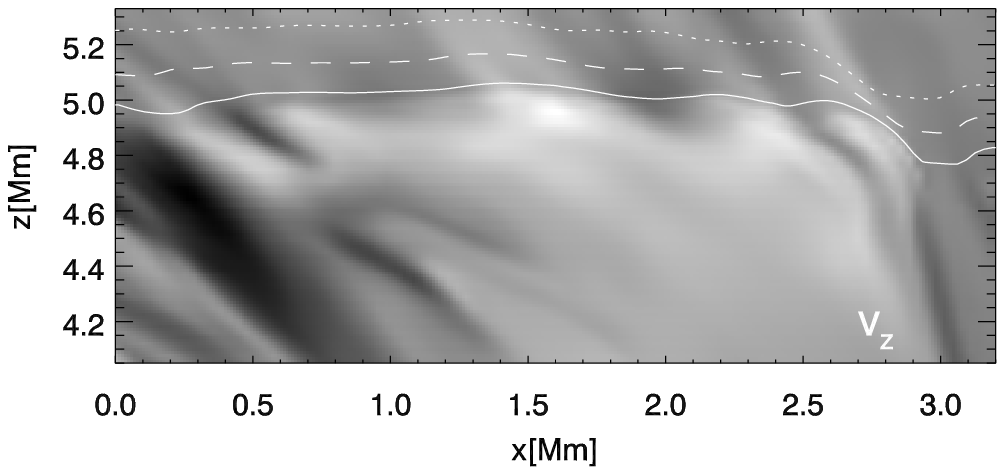}}}
  \caption{Vertical cuts along the center of the filament 
   at constant $y=0.9\,$arcsec in Fig.~\ref{fig:filament_hor}.
   The grey-scale ranges between
   minimum (black) and maximum (white) of the
   quantities shown are: $|B|$: 480$\,$G ... 4220$\,$G; $B$ inclination
   with respect to vertical: 20.1$\,$deg ... 90$\,$deg; $-v_x$:
   $-1.3\,$km$\cdot$s$^{-1}$ ...  3.6$\,$km$\cdot$s$^{-1}$; $v_z$:
   $-2.7\,$km$\cdot$s$^{-1}$ ...  2.5$\,$km$\cdot$s$^{-1}$.}
  \label{fig:filament_vertz}
\end{figure*}

\begin{figure}
  \centering 
  \resizebox{\hsize}{!}{\includegraphics{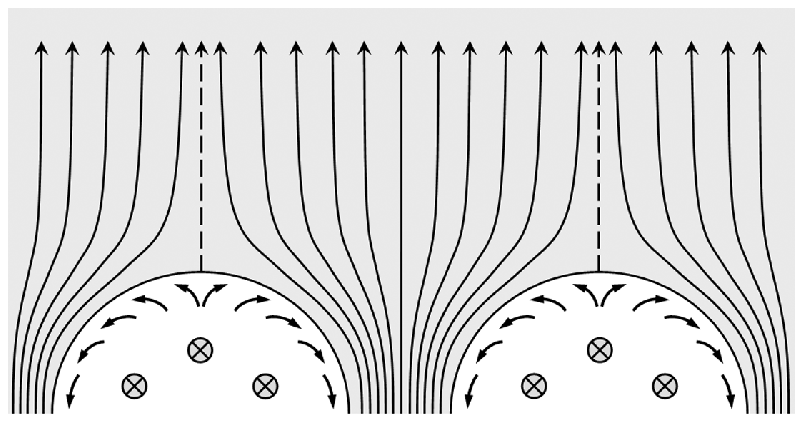}}
  \caption{Schematic illustration of the basic structure of penumbral
  filaments as suggested by \citet{Zakharov:etal:2008} on the basis of
  spectro-polarimetric observations. Shown is a vertical cut
  perpendicular to the filament axes. The bright semicircular areas
  indicate the uppermost part of the upflow plume, the curved arrows
  represent the overturning convective flow. Circled crosses indicate the
  almost horizontal magnetic field and velocity along the filaments. The
  lines in the grey area are projections of the less inclined magnetic
  field lines outside the filaments. The intrusion of the plume and flux
  transport by the overturning motion have pushed aside the less inclined
  field between the filaments.}
  \label{fig:sketch}
\end{figure}

For the detailed analysis of the simulation results, we consider a
snapshot taken $30$ solar minutes after the start of the high-resolution
run. We first present results concerning the global structure of the
simulated spot and its average properties and then consider the
penumbral filaments.
 
\subsection{Large-scale structure}

Fig.~\ref{fig:intensity_global} presents a continuum intensity image
($\lambda=630\,$nm) of the sunspot and its environment. For better
visibility, the snapshot has been doubled in $y$-direction.  The spot
umbra has a width of about $10$ Mm and is surrounded on both sides by a
penumbra of about $4-5$ Mm width, harboring filaments with dark
cores. The umbra shows umbral dots similar to those found by
\citet{Schuessler:Voegler:2006} in a local simulation.

Fig.~\ref{fig:bb_vert_global} displays the square root of the magnetic
field strength on a vertical cut through the simulation box at
$y=3.84$~Mm. The height expansion of the sunspot due to the decreasing
gas pressure is well visible. Comparison with
Fig.~\ref{fig:intensity_global} shows that the regions corresponding to
the penumbra (roughly in the ranges $x=9...14\,$Mm and
$x=23...28\,$Mm) mostly have a deep underlying magnetic structure. In
fact, about the same amount of vertical flux emerges in the penumbral
part of the simulated spot as in the umbra, so that the penumbra cannot
be shallow \citep[cf.][]{Solanki:Schmidt:1993}.

Fig.~\ref{fig:bb_vert_global} also indicates that the magnetic field is
rather inhomogeneous between the level $\tau_{630}=1$ (indicated by the
white line) and about 2~Mm below. This is caused by magneto-convection
in the form of upflow plumes which lead to reduced field strength owing
to expansion and flux transport by the overturning flow. As we shall
discuss further below, the underlying processes for the formation of
penumbral filaments are the same as those for the umbral dots
\citep{Schuessler:Voegler:2006}. Above $\tau_{630}=1$, the field again
becomes rather smooth: owing to the low plasma $\beta$ in these layers
overlying umbra and penumbra, gas pressure gradients cannot maintain
strong field inhomogeneities and the field structure has to become
almost force-free. 

The sunspot has already lost some amount of flux to its environment,
which shows a plage-like mean vertical flux density of about 200~G at
$\tau_{630}=1$. Most of this flux has been assembled in small-scale flux
concentrations in the intergranular lanes (bright structures in
Fig.~\ref{fig:intensity_global}), but also pore-like structures have
formed, probably supported by the periodic boundary condition.

A more quantitative account of the large-scale structure of the
simulated sunspot is provided by Fig.~\ref{fig:means_global}, which
shows horizontal profiles of average quantities determined in the layers
accessible to observation. The profiles of the magnetic field components
and its inclination with respect to the vertical agree well with the
observational curves determined by \citet{Keppens:Martinez:1996},
although the field strength in the simulated penumbra is somewhat
high. Average outward horizontal flows with peak velocities of about
$2\,$km$\cdot$s$^{-1}$ are clearly visible in both penumbral
regions. The main discrepancy between the observed and simulated
penumbra is the lower mean brightness in the simulation (about 60\% of
the quiet Sun) compared to the observed value of about 75\%
\citep[e.g.,][]{Schlichenmaier:Solanki:2003}. Guided by our experience
with umbra simulations, we conjecture that this discrepancy could be due
to insufficient thermal relaxation of the magnetic region, so that the
magneto-convective processes providing the structuring and the energy
transport in the penumbra are still not completely developed. 

\subsection{Penumbral filaments}
Fig.~\ref{fig:intensity_detail} shows an enlargement of the penumbral
region on the right-hand side of the continuum intensity map
(Fig.~\ref{fig:intensity_global}). The most conspicuous structures are
dark-cored filaments of up to a few Mm length. On their end facing the
umbra, the filaments typically show bright `heads', which propagate into
the umbra. The typical lifetime of the filaments is about one hour. The
three lines in Fig.~\ref{fig:intensity_detail} indicate the positions of
vertical cuts shown in Fig.~\ref{fig:cuts_color}, which give the
magnetic field strength, $B$, the inclination angle with respect to
the vertical, ${\arcsin}(B_x/\vert B\vert)$, and the horizontal flow
velocity, $v_x$.

It is obvious from Fig.~\ref{fig:cuts_color} that the penumbral
filaments correspond to slender structures of about $2$ Mm depth (near
the umbra) and a few hundred km width with significantly reduced overall
field strength. The vertical field component, $B_z$, is reduced to close
to zero near the top of the filament, so that an inclination angle of
almost $90\deg$ results, i.e., the field becomes nearly horizontal. In
the inner penumbra, the filaments develop within regions of strong
magnetic field; only in the periphery of the spot, where the vertical
thickness of the penumbra drops to less than $2$ Mm, the filaments
become more connected to the almost field-free convection zone
beneath. Therefore, the filaments in this simulation do not originate
from convection penetrating into the penumbra from beneath; they are
rather similar to the plume structures leading to umbral dots in the
simulation of \citet{Schuessler:Voegler:2006}, here modified by the
presence of a significant horizontal field component. 

The third column in Fig.~\ref{fig:cuts_color} shows that the penumbral
filaments are associated with strong horizontal flows away from the
umbra. These flows reach their largest velocities of a few
km$\cdot$s$^{-1}$ in the upper and outer parts of the filaments, where
the magnetic field is most strongly inclined.

Fig.~\ref{fig:profiles} provides a detailed view of the physical
conditions in the filaments by showing horizontal and vertical profiles
of magnetic field and velocity components for the cut shown in the
middle row of Fig.~\ref{fig:cuts_color}. The horizontal profiles at
$z=5$ Mm in the top panel illustrate the connection between horizontal
outflow and almost horizontal magnetic field. The vertical cuts at the
positions $y=1.94$ Mm and $y=2.6$ Mm, respectively, shown in the middle
and bottom panels of Fig.~\ref{fig:profiles} indicate that the outflows
are concentrated in the near-surface layers.  It is tempting to
associate these flows with the Evershed effect
\citep[cf.][]{Scharmer:etal:2008}.

In what follows, we will study in more detail one prototypical filament
(outlined by the white square in Fig.~\ref{fig:intensity_global}).
Fig.~\ref{fig:filament_hor} shows a magnified continuum intensity image
at 630~nm (bottom left panel) together with maps of various quantities
calculated at the surface $\tau_{630}=0.1$, roughly corresponding to the
layer dominating the spectro-polarimetric information obtainable with
the often used neutral iron lines at 630.15~nm and 630.25~nm. This gives
a first idea of the actually observable consequences of the physical
processes underlying the penumbral filamentation, although definite
results will require a detailed comparison with synthetic Stokes
profiles from a non-grey simulation, also taking into account image
degradation by a realistic point-spread function and noise. Such a study
is beyond the scope of this paper.

The intensity image in Fig.~\ref{fig:filament_hor} shows a bright
filament of a few hundred km width, pervaded by a dark lane of about
100~km width. The head of the filament has moved some way into the umbra
and has nearly disconnected from the filament, forming a peripheral
umbral dot. The maps at constant optical depth $\tau_{630}=0.1$ show an
upflow ($v_z>0$) of up to 1.5 km$\cdot$s$^{-1}$ at the center of the
filament, which is connected to downflows at both sides via outflows in
$\pm y$-direction, perpendicular to the filament axis. In combination
with the longitudinal flow along the filament, such a flow pattern is
consistent with the recent observations of `twisting motions' in
penumbral filaments \citep{Ichimoto:etal:2007,Zakharov:etal:2008}.  The
strong horizontal flow away from the umbra ($-v_x>0$) along the dark
lane is associated with a weaker, laterally more extended, inward return
flow at the periphery of the filament. The profiles of the three
velocity components along a cut in $y$-direction through the filament at
$x=2\,$arcsec shown in the upper panel of Fig.~\ref{fig:B_v_profiles}
clearly reveal the flow pattern of a rising plume with laterally
overturning motion and a strong longitudinal outflow.

The maps of the magnetic field components in Fig.~\ref{fig:filament_hor}
and the corresponding profiles along the perpendicular cut at
$x=2\,$arcsec (lower panel of Fig.~\ref{fig:B_v_profiles}) show a
reduction of the vertical component, $B_z$, while the component along
the filament, $B_x$, stays almost constant. Accordingly, the field is
more strongly inclined and its strength reduced in the filament.  The
component perpendicular to the filament axis reverses sign at the
filament center (above the dark lane), corresponding to the cusp-shaped
geometry of the top part of the filament: the strong field sideways of
the filament expands and `closes' above the filament. This configuration
has recently been confirmed observationally by
\citet{Borrero:etal:2008}.

The structure below the visible filament and its relationship to the
surfaces of constant optical depth that are relevant for observations
are illustrated by vertical cuts perpendicular to and along the
filament, which are shown in Figs.~\ref{fig:filament_vertx} and
\ref{fig:filament_vertz}, respectively. The perpendicular cuts at
$x=2\,$arcsec (Fig.~\ref{fig:filament_vertx}) reveal a structure which
is very similar to that underlying the umbral dots in the simulations of
\citet{Schuessler:Voegler:2006}: a strong upflow plume in the center
with narrow downflows at its sides leads to an elevation of the
iso-$\tau$ surfaces by about 200~km, also clearly visible in the
temperature profile. The magnetic field is strongly reduced in the plume
since 1) plasma moving upward expands owing to the strong pressure
stratification and 2) the overturning horizontal flow transports flux to
the downflow regions at the edges of the filament. The latter effect
preferentially weakens the vertical field component while horizontal
field is replenished by the upflow. The presence of the strongly
inclined magnetic field deflects the central upflow outward, leading to
a horizontal flow of about $2\,$km$\cdot$s$^{-1}$ away from the umbra of
the spot.

There is some indication that occasionally part of the overturning flow
is recirculated into the upflow plume, so that the flow pattern becomes
reminiscent to roll convection as originally suggested by
\citet{Danielson:1961}. To see this, consider the velocity components
shown in the lower panels of Fig.~\ref{fig:filament_vertx}: the central
upflow ($v_z$) drives perpendicular outflows ($v_y$) near the top of the
filament, which feed the downflows adjacent to it. About 300~km deeper,
the sign of $v_y$ is reversed, so that we now have a perpendicular
inflow, which converges with the central upflow and closes the roll. It
is not clear at this moment 1) whether this is a robust or a transient
feature of the flow pattern, 2) whether it is really roll-type
convection or more related to Kelvin-Helmholtz instability of the
downflow, and 3) whether it is realistic at all since the simulation
still has a much larger effective magnetic diffusivity than the plasma
in a real sunspot. Anyway, its rather small depth extension indicates
that such roll convection probably does not contribute much to the convective
energy transport, which is mainly provided by the deep upflow plume.

Fig.~\ref{fig:filament_vertz} shows physical quantities on a vertical
cut roughly along the dark lane. It reveals that the structure
underlying the visible filament is extended in depth as far as the
upflows and the reduction of the field strength are concerned.  On the
other hand, the horizontal outflow ($v_x$) and the field inclination
sharply peak near optical depth unity and lead to the formation of a
narrow, almost horizontal flow channel.

Altogether, the structure of the simulated penumbral filaments may well
be represented by the sketch shown in Fig.~\ref{fig:sketch}, which
illustrates the essence of the high-resolution spectro-polarimetric
observations of \citet{Zakharov:etal:2008}.

A direct connection of the Evershed flow with the basic
magneto-convective process in the penumbra has been suggested by
\citet{Scharmer:etal:2008} on the basis of the simulations of
\citet{Heinemann:etal:2007}.  In fact, we find a dominant outward flow
due to the deflection of the upward flow in the central part of the
filament, which is turned outward by the presence of the inclined
magnetic field. The corresponding return flow is mainly located in the
regions with less inclined and stronger magnetic field between the
filaments (see Figs.~\ref{fig:filament_hor} and \ref{fig:B_v_profiles});
it therefore has a smaller horizontal component at the same (optical)
depth. Consequently, the observable average horizontal velocity is
outward (see the lower panel of Fig.~\ref{fig:means_global}) in
accordance with the observed Evershed effect, even though there is no
significant overall net outward mass flux in the penumbra. This possibly
resolves the long-standing problem of the source and disposal of the
mass transported by the Evershed flow
\citep[e.g.,][]{Solanki:etal:1994}. The existence of the weaker inward
return flows is a prediction of the simulation that should be testable
with high-resolution observations. Note that \citet{Ichimoto:etal:2008}
have found first indications for flows between the Evershed flow
channels. In our current simulation, the penumbral filaments appear to
be more separated from each other than shown by observations. While the
return flow is clearly visible around separated individual filaments,
there are indications that it might be less easily detectable in the
case of more densely packed filaments. This needs to be investigated in
the future with simulations carried out at higher spatial resolution.

The hot upflow extends over most of the length of the filaments, which
provides a rather efficient energy transport to supply the radiative
losses of the penumbra. In contrast, the moving-flux-tube model with
only localized upflows is far too inefficient to explain the average
brightness of the penumbra
\citep{Schlichenmaier:Solanki:2003}. Similarly, the localized plumes
underlying umbral dots can only maintain the much lower brightness of
the umbra in comparison with the penumbra, although the basic
magneto-convective process is very similar.

The location of the iso-$\tau$ lines in Figs.~\ref{fig:filament_vertx}
and \ref{fig:filament_vertz} demonstrates that the main part of the
upflow plume and the overturning motion underlying the filament is below
the visible layers. Spectro-polarimetric observations corresponding to
$\tau_{630}\simeq 0.1\dots0.01$ just scratch the `tip of the iceberg'
and reveal only the uppermost part of the Evershed flow and magnetic
cusp structure, which is already laterally much more homogeneous than
the underlying main part of the filament.  This situation is very
similar to that of umbral dot observations and in the past probably has
led, together with the effect of insufficient spatial resolution, to
ambiguities in the interpretation of spectroscopic observations of
velocity and magnetic field.

\section{Discussion}

The properties of our simulated sunspot are consistent with the general
picture of sunspot structure that has emerged from observational
studies. This applies to the overall structure (e.g., distinction
between umbra and penumbra, average `radial' profiles of the magnetic
field components and inclination angle, average outflow in the penumbra)
as well as to the detailed properties of the fine structure of umbra and
penumbra. The penumbral filamentation results from magneto-convective
energy transport in the form of hot rising plumes, very similar to the
process giving rise to umbral dots \citep{Schuessler:Voegler:2006}. The
inclined magnetic field near the periphery of the spot causes a symmetry
breaking which leads to elongated filaments with strong outflows along
flow tubes of nearly horizontal field near optical depth unity. In
addition to the flow along the filament, the upflow also turns over into
a motion perpendicular to the filament axis. Dark lanes appear above the
strongest upflows owing to the upward bulging of the surface of optical
depth unity and the piling up of plasma in a cusp-shaped region at the
top of the filament, above which the less inclined field outside the
filament becomes laterally fairly homogeneous. The horizontal outflows
are concentrated along the dark lanes. All these properties are
consistent with recent observational results
\citep[e.g.,][]{Bellot-Robio:etal:2005, Rimmele:Marino:2006,
Langhans:etal:2007, Ichimoto:etal:2007, Borrero:etal:2008,Noort:Rouppe:2008,
Zakharov:etal:2008}.

Our results are also consistent with many properties of the short
penumbral filaments found by \citet{Heinemann:etal:2007}, who gave
interpretations along very similar lines. The fact that simulations with
two rather different numerical codes lead to basically the same picture
for the formation of penumbral structure indicates that, in spite of all
differences in detail, the simulations have indeed captured 
essential physical processes.  The explanation for the Evershed effect
as a natural consequence of rising plumes in an inclined field
\citep[cf.][]{Scharmer:etal:2008} connects this flow directly to the
basic magneto-convective structure of the penumbra, so that it should
occur whenever penumbral structure is present.  The geometry of the flow
pattern is such that the observable average outflow velocity need not to
be connected with a net outflux of mass. This does not exclude the
additional presence of siphon flows \citep[e.g.,][]{Degenhardt:1993,
Montesinos:Thomas:1997, Solanki:etal:1994}, but it is much less clear
whether the pressure gradients required for a sustained outward siphon
flow are maintained always and everywhere in all penumbrae.

What can be said on the basis of our simulation results about the
various models that have been proposed to explain the penumbral
structure? First of all, we do not see evidence for the `moving flux
tube' model and interchange convection
\citep[e.g.][]{Schlichenmaier:etal:1998b, Schlichenmaier:etal:1998a}.
In our simulations, the progression of the filament heads toward the
umbra during their formation phase is not caused by the inward motion of
a narrow flux tube, but rather due to the expansion of the sheet-like
upflow plumes along the filament.

Furthermore, the simulations combine various aspects of the `embedded
flux tube' model \citep{Solanki:Montavon:1993, Borrero:etal:2005,
Borrero:etal:2006} and the `gappy penumbra' model
\citep{Spruit:Scharmer:2006,Scharmer:Spruit:2006}. However, the
simulated penumbral filaments are neither intrusions of field-free
plasma from below nor are they confined to almost horizontal flux tubes
disconnected from their environment, particularly in depth. The
uppermost part of the plume structure forming the penumbral filament
with its strong horizontal flow and almost horizontal field could
possibly be represented by a kind of embedded flux tube. The main
feature missing in the embedded flux tube models is the overturning
convection within the filament and the deep-reaching upflow of plasma
that provides the primary energy supply. In this respect, the underlying
plume structure with its reduced (albeit non-vanishing) field strength
has much similarity with the `gappy' configuration of the penumbra. This
scenario captures the convective origin of the penumbral filamentation,
even though the gaps form within the strong magnetic field. As a
consequence, the gaps contain a horizontal field and, in most cases, are
not connected to the almost field free convecting plasma below the
penumbra.

There is no clear evidence in our simulations that the penumbral
structure is affected or even caused by the fluting instability as
suggested by \citet{Weiss:etal:2004}. However, the periodic boundary
condition in the horizontal ($x$) direction used in the simulation
implicitly corresponds to the existence of identical sunspots just
about 20~Mm from the penumbral boundaries, which certainly affects the
field structure, particularly the inclination, in the outer
penumbra. Therefore, the simulation possibly does not well represent the
convective pumping effect suggested by \citet{Weiss:etal:2004} as a
mechanism for the downward dragging of magnetic flux in the outer
penumbra. This might provide a possible explanation for the still rather
small extension of the simulated penumbra. Observations in fact indicate
that penumbrae are often suppressed on the side of a sunspot which faces
a nearby spot of the same polarity.

Altogether, our results indicate a new level of realism in the
theoretical modelling of sunspot structure. The properties of the
simulated penumbral filaments are consistent with a variety of
observational results and provide a basis for a physical understanding
of umbral and penumbral structure in terms of magneto-convective
processes. On the other hand, there are still clear discrepancies
between the numerical results and real sunspots, so that there is some
way ahead to be covered towards a completely satisfactory model. Our
penumbral structure does not yet appear to be fully evolved and the
overall extension of the penumbra is still somewhat small. The average
intensity profile indicates that we have simulated the development of an
inner penumbra, while the outer penumbra might be more strongly affected
by convective pumping \citep{Weiss:etal:2004}. 

The lower boundary condition remains arbitrary since we still have no
reliable observational constraints concerning the subsurface structure
of sunspots.  Computational limits have forced us to use a rather coarse
spatial resolution of 32~km and a fairly small computational
box. Furthermore, we could only cover a relatively short overall
evolution time. As a consequence, the effective diffusivities in the
simulation are still much larger than the real values. Test calculations
with different resolution show that 1) first indications for filamentary
structure appear already at a horizontal resolution of 96~km and 2) the
reduction of field strength in the plumes increases somewhat when we
move to a resolution of 24~km.  On that basis, the fundamental physical
process of sheet-like plume convection appears to be a robust
feature. The results will certainly change in detail (and, hopefully,
become even more similar to the observed penumbrae) as resolution
increases, but we do not expect totally new processes replacing those
that we have described here.

The rapid increase in available computational power and the foreseeable
progress in local helioseismology will soon alleviate some of the
limitations of the present approach and thus enable us to carry out even
more realistic simulations. 

\acknowledgements
Vasily Zakharov kindly provided Fig.~\ref{fig:sketch}.
M. Rempel wishes to thank the Institute for Pure and Applied Mathematics
(IPAM) of UCLA, Los Angeles for their support to attend the program on
`Grand Challenge Problems in Computational Astrophysics', from which
this work has benefited significantly. M. Rempel also thanks J.~M. Borrero
for in-depth discussions regarding observations and models of sunspot
structure. The National Center for Atmospheric Research (NCAR) is sponsored 
by the National Science Foundation.

\appendix
\section{Numerical scheme}
\label{appendix}
To alleviate the time step constraint for an explicit code brought about
by high Alfv{\'e}n velocity, we limit the strength of the Lorentz
force in the regions of low $\beta$, such that the resulting Alfv{\'e}n
velocity has a given upper bound, $c_{\rm max}$:
\begin{equation}
  \vec{F}_L=\frac{c_{\rm max}^2}{\sqrt{c_{\rm max}^4+v_{\rm A}^4}}\,
    \vec{J}\times\vec{B}\;.
\end{equation}
Here $v_{\rm A}$ denotes the value that the Alfv{\'e}n
velocity would have without limitation. The modification of the Lorentz
force is applied to regions where $v_{\rm A} > c_{\rm max}$
or, in terms of the $\beta$ value, where
\begin{equation}
  \beta<\frac{8\pi\,p}{\varrho c_{\rm max}^2}\approx 
  \left(\frac{c_{\rm sound}}{c_{\rm max}}\right)^2 \approx 0.05\;.
\end{equation}
For the simulations presented in this paper, we have used a value of
$c_{\rm sound}=7$ km$\cdot$s$^{-1}$ to estimate the speed of sound above
the sunspot umbra and set $c_{\rm max}=31$ km$\cdot$s$^{-1}$.  Since the
magnetic field is already in an almost force free state for
$\beta<0.05$, this correction has only a minor influence on the overall
force balance. We have tested this with a 2D run for which we varied the
value of $c_{\rm max}$ by a factor of $3$ and found no significant
difference. This approach increases the explicit time step limit by
about $2$ orders of magnitude at zero computational expense, in contrast
to the alternative of an implicit treatment of the Lorentz force.

The presence of high and low $\beta$ regions as well as high and low
Mach number flows presents a significant challenge for a numerical
scheme to properly resolve all regimes without being too diffusive in
any of them.  To this end, we have changed the artificial diffusivity
scheme of the {\em MURaM} code. After a piecewise linear reconstruction of the
solution, $u_i$, where the reconstruction slope in each cell $\Delta
u_i$ is limited by an appropriate slope limiter, we use the extrapolated
values at the interface $u_l=u_i+0.5\,\Delta u_i$ and
$u_r=u_{i+1}-0.5\,\Delta u_{i+1}$ to compute the diffusive flux
\begin{equation}
  F_{i+\frac{1}{2}}=\frac{1}{2}\, c_{i+\frac{1}{2}}\,
  \phi(u_r-u_l,u_{i+1}-u_{i})\,\left(u_r-u_l\right)\;,
\end{equation}
where $c$ denotes the characteristic velocity. Using $\phi=1$ reduces
the scheme to a standard (at least) second order flux (depending on
limiter) such as used in a second-order Lax Friedrichs scheme. For the
$4$th order MHD scheme of the {\em MURaM} code, we found that a choice of
$\phi=\left(\frac{u_r-u_l}{u_{i+1}-u_{i}}\right)^2$ for
$(u_r-u_l)\cdot(u_{i+1}-u_{i})>0$ and $\phi=0$ otherwise (no artificial
steepening) represents the best compromise between maximum stability and
minimum diffusion. The diffusivity of the scheme is controlled through
the slope limiter, for which we use a linear combination of the most
diffusive Minmod and least diffusive Superbee limiter 
\citep[see, e.g.][]{LeVeque:1990}:
$\eps\,{\rm Minmod}+(1-\eps)\,{\rm Superbee}$ and allow $\eps$ to vary
as function of $\beta$ (typically $0.5$ in high- and less than $0.2$ in
low-$\beta$ regions).  Furthermore it turned out that it is sufficient
to use $c=0.1\,c_{\rm sound}+v+v_{\rm alf}$ for the characteristic
velocity, which significantly reduces the diffusivity in low Mach number
flows. We apply this scheme to all MHD variables and account for effects
of (unfortunately unavoidable) mass diffusion in the momentum and energy
fluxes. The div $\vec{B}$ error produced by the diffusion scheme is
controlled by iterating
\begin{equation}
  \frac{\partial \vec{B}}{\partial t}=\mu (\Delta x)^2
       {\rm grad} ({\rm div}\vec{B}) \;.
\end{equation}
For values of $\mu=0.3\ldots 0.5$, typically less than $10$ iterations
are required to satisfy ${\rm max}\,(\Delta x\,\vert{\rm
div}\vec{B}\vert\,) < 10^{-3} B_{\rm rms}$, which was found to be
sufficient. In the results presented in this paper, we did not use any
explicit viscosity or magnetic diffusivity. The artificial diffusivities
are added once every time step, after completion of the Runge-Kutta loop
of the $4^{\rm th}$ order MHD scheme.  The treatment of diffusivity
outlined above leads to a scheme that is fully shock-capturing, at least
$4$th order accurate in smooth regions (higher order is possible
depending on slope limiter used), minimally diffusive for low Mach
number flows, and stable to $\beta$ values as low as $10^{-4}$.

A very low value of $\beta$ also leads to problems in codes that use the
conservative formulation of the energy equation, since the determination
of the internal energy requires to compute the small difference between
the nearly equal values of the total and the magnetic energies.  To
avoid this problem, we switch to an isothermal equation of state in
regions where $E_{\rm int}<10^{-3}E_{\rm mag}$ and prevent too small
$\beta$ values by imposing a lower limit for the density when $E_{\rm
int}<10^{-5}E_{\rm mag}$. An alternative to this procedure would have
been to directly solve an equation for the internal energy at the
expense of loosing the advantage of the conservative formulation. Since
most of the dynamics are driven beneath the photosphere and stability
problems only occur in layers with an average density at least $3$
orders of magnitude less than the photosphere (and therefore of little
dynamical importance), we have preferred to stay with the conservative
formulation.

\end{document}